\documentstyle[12pt,twoside]{article}

\renewcommand{\baselinestretch}{1.25} 
\makeatother

\catcode `\@=11
\@addtoreset{equation}{section}

\title{ \LARGE  \bf Towards the Born-Weyl Quantization of Fields\thanks{to 
appear in {\em International Journal of Theoretical Physics, }  January 1998}
{\it  }
}
\author{ Igor V. Kanatchikov\thanks{
Dept. of Analytical Mechanis and Field Theory, 
Institute of Fundamental Technological Research, 
Polish Academy of Sciences,  
Swietokrzyska 21, 00-049 Warsaw, Poland.  
E-mail: ikanat@ippt.gov.pl   }
}
\date{}

\begin{document}

\maketitle




\renewcommand{\@oddhead}{{\footnotesize\sf Born-Weyl Quantization of Fields } 
\hfil \bf\thepage}
\renewcommand{\@evenhead}{ \bf\thepage \hfil  { \footnotesize\sf Kanatchikov} }

\newcommand{\beq}{\begin{equation}}
\newcommand{\eeq}{\end{equation}}
\newcommand{\beqa}{\begin{eqnarray}}
\newcommand{\eeqa}{\end{eqnarray}}
\newcommand{\nn}{\nonumber}
 
\newcommand{\half}{\frac{1}{2}}
 
\newcommand{\xt}{\tilde{X}}
 
\newcommand{\uind}[2]{^{#1_1 \, ... \, #1_{#2}} }
\newcommand{\lind}[2]{_{#1_1 \, ... \, #1_{#2}} }
 
\newcommand{\com}[2]{[#1,#2]_{-}} 
\newcommand{\acom}[2]{[#1,#2]_{+}} 
\newcommand{\compm}[2]{[#1,#2]_{\pm}}
 
\newcommand{\lie}[1]{\pounds_{#1}}
\newcommand{\co}{\circ}
\newcommand{\sgn}[1]{(-)^{#1}}
\newcommand{\lbr}[2]{ [ \hspace*{-1.5pt} [ #1 , #2 ] \hspace*{-1.5pt} ] }
\newcommand{\lbrpm}[2]{ [ \hspace*{-1.5pt} [ #1 , #2 ] \hspace*{-1.5pt}
 ]_{\pm} }
\newcommand{\lbrp}[2]{ [ \hspace*{-1.5pt} [ #1 , #2 ] \hspace*{-1.5pt} ]_+ }
\newcommand{\lbrm}[2]{ [ \hspace*{-1.5pt} [ #1 , #2 ] \hspace*{-1.5pt} ]_- }
\newcommand{\pbr}[2]{ \{ \hspace*{-2.2pt} [ #1 , #2 ] \hspace*{-2.55pt} \} }
\newcommand{\we}{\wedge}
\newcommand{\dv}{d^V}
\newcommand{\nbrpq}[2]{\nbr{\xxi{#1}{1}}{\xxi{#2}{2}}}
\newcommand{\lieni}[2]{$\pounds$${}_{\stackrel{#1}{X}_{#2}}$  }

\newcommand{\rbox}[2]{\raisebox{#1}{#2}}
\newcommand{\xx}[1]{\raisebox{1pt}{$\stackrel{#1}{X}$}}
\newcommand{\xxi}[2]{\raisebox{1pt}{$\stackrel{#1}{X}$$_{#2}$}}
\newcommand{\ff}[1]{\raisebox{1pt}{$\stackrel{#1}{F}$}}
\newcommand{\dd}[1]{\raisebox{1pt}{$\stackrel{#1}{D}$}}

\newcommand{\nbr}[2]{{\bf[}#1 , #2{\bf ]}}
\newcommand{\der}{\partial}

\newcommand{\oo}{$\Omega$}
\newcommand{\Om}{\Omega}
\newcommand{\om}{\omega}
\newcommand{\eps}{\epsilon}
\newcommand{\si}{\sigma}
\newcommand{\Lm}{\Lambda^*}
\newcommand{\ga}{\gamma} 

\newcommand{\inn}{\hspace*{2pt}\raisebox{-1pt}{\rule{6pt}{.3pt}\hspace*
{0pt}\rule{.3pt}{8pt}\hspace*{3pt}}}
\newcommand{\sro}{Schr\"{o}dinger\ }
\newcommand{\bm}{\boldmath}

\newcommand{\vol}{\omega}
\newcommand{\dvol}[1]{\der_{#1}\inn \vol}
 
\newcommand{\bd}{\mbox{\bm $d$}}
\newcommand{\bder}{\mbox{\bm $\der$}}
\newcommand{\bI}{\mbox{\bm $I$}}

\newcommand{\dxo}[2]{dx^{#1_1}\otimes ... 
\otimes dx^{#1_{#2}}} 
\newcommand{\dero}[2]{\der_{#1_1}\otimes ... 
\otimes \der_{#1_{#2}}} 
 
\newcommand{\lapl}{\bigtriangleup}
\newcommand{\psib}{\bar{\psi}}
\newcommand{\derts}{\stackrel{\leftrightarrow}{\der}}
\newcommand{\what}[1]{\widehat{#1}}


\renewcommand{\mu}{i}
\renewcommand{\nu}{j}
\newcommand{\gmu}{\gamma^\mu}
\newcommand{\gnu}{\gamma^\nu}
\newcommand{\ka}{\kappa}
\newcommand{\hka}{\hbar \kappa}
\newcommand{\al}{\alpha}

\newcommand{\bx}{{\bf x}}
\newcommand{\bk}{{\bf k}}
\newcommand{\omk}{\omega_{\bf k}}


\begin{quote}
--------------------------------------------------------------------------------------------------- \\
{\small 
Elements of the quantization in field theory based on the 
covariant polymomentum Hamil\-tonian formalism 
	(the De Donder-Weyl theory),   
a  possibility of which was originally discussed in 1934 by 
Born and Weyl, are developed. 
The approach is based on a recently proposed 
graded 
Poisson bracket on 
differential forms in field theory.  
A covariant analogue of the Schr\"odinger 
equation for a hypercomplex wave function is put forward. 
It leads to the De Donder-Weyl Hamilton-Jacobi equations in 
quasiclassical limit. 
A possible relation to the functional Schr\"odinger picture 
in quantum field theory is outlined.} \\ 
---------------------------------------------------------------------------------------------------
\end{quote}

\medskip

\section{\large \bf
INTRODUCTION}
 
\large 

The approach to the canonical quantization of field theories 
which 
was originally developed 
shortly after 
the formalism of quantum mechanics was established 
is based on the 
representation of fields as mechanical systems with an infinite 
number of degrees of freedom. This approach was essentially inspired 
by what 
Heisenberg and Pauli
referred to in their private correspondence as  ``Volterra 
Mathematik'' namely, some developments of that time 
in the functional and variational 
calculus (cf. e.g. (Volterra, 1959)).
The canonical quantization of fields commonly known since then  is 
based on the only generalization of the Hamiltonian formalism to  
field theory  available at that time.  
However, shortly thereafter in the papers by Carath\'eodory,
De Donder, Weyl and others 
(see e.g. (Rund, 1966) for a review)  
on the calculus of variations of multiple integrals  
different alternative ways of extending  
the Hamiltonian formulation to field theory have appeared   
which where unified within a general scheme later in the 
forties by Lepage 
(see e.g. (Kastrup, 1983) for a review and references).  
Unlike the standard Hamiltonian formalism in field theory 
all these formulations do not 
distinguish between the space and time coordinates 
and do not refer to an infinite dimensional phase space.  
Instead, fields are treated rather as 
a sort of 
dynamical systems with several ``times'' the 
role of which is played by all space-time coordinates on equal footing.   
In doing so the phase space is replaced by what is called below 
the {\em polymomentum phase space}, a finite dimensional space of 
field variables and ``polymomenta'' 
which are defined from the Lagrangian as  the 
conjugate  momenta associated with 
each space-time derivative of the field (see Sect. 2). 
In the case of 
one-dimensional "space-time" this picture
reproduces the standard Hamiltonian formalism in mechanics 
 which underlies the canonical quantization. 
It is quite natural 
to ask whether the above mentioned "polymomentum" Hamiltonian formulations  
can provide us with a basis for a quantization procedure in field theory. 
A priori the manifest space-time symmetry 
	of these formulations 
and the finite dimensionality of the 
polymomentum phase space can be viewed as potential advantages of such 
an approach which may look especially appropriate in the context of 
quantization of  General Relativity. 

For the first time the problem of field quantization based on 
a polymomentum formulation was discussed by 
Born (Born, 1934) and Weyl (Weyl, 1934). 
I, therefore,  refer to the corresponding program 
as the Born--Weyl quantization. 
Unfortunately,  
although the question was not clarified, there were essentially 
no  discussions of the issue  since then and only a few references 
touching this problem  could be cited 
(see e.g.   quotations in (Kanatchikov, 1996)).  
The main reason for this   
(besides possible historical ones) 
seems to be the lack of an appropriate 
generalization of Poisson brackets  
to the framework of the polymomentum formulations.  
However, recently such 
brackets 
were constructed within the  De Donder--Weyl canonical theory, 
the simplest representative of the Lepagean canonical theories 
(see  (Kanatchikov, 1996)). 
Elements of this construction 
are briefly described in Sect. 2. 
The purpose of the present communication is 
to discuss a possible approach to field quantization based on these 
brackets.  



\section{\large \bf 
DE \, DONDER--WEYL \, FORMULATION \, 
AND\,  THE \, \\
POISSON \, BRACKETS\,  ON \, FORMS}

For the first order Lagrangian field theory 
	given by the Lagrangian density 
$L=L(y^a,\der_i y^a, x^i)$, where 
$x^i$ $ (i=1,...,n)$ are    space-time 
coordinates and $y^a$ $ (a=1,...,m)$ are  field variables, 
let us define the {\em polymomenta} $p_a^i$ 
and the {\em De Donder--Weyl (DW) Hamiltonian function} $H$:
\beq
p^i_a := \der L /\der(\der_i y^a), \quad  H := p^i_a\der_iy^a -L .
\eeq
Then the second order Euler-Lagrange equations can be rewritten in 
the following first order form (see for instance (Rund, 1966)) 
\beq
\der p^i_a / \der x^i=- \der H /\der y^a , 
\quad 
\der y^a / \der x^i=\der H  / \der p^i_a  
\eeq
which reproduces Hamilton's canonical equations in mechanics when $n=1$. 
Therefore eqs. (2.2) can be viewed as a covariant generalization 
of the Hamiltonian formulation to field theory,
to be  referred to as  
the {\em DW Hamiltonian formulation} 
in the following.
An interesting question is how other elements of the standard 
Hamiltonian formalism in mechanics can be extended to the present 
polymomentum formulation of field theory. 
\nopagebreak

An essential ingredient of the canonical formalism is the 
Hamilton-Jacobi  
 (HJ) theory which also has its counterpart  
within the DW  formulation. The
corresponding DW HJ equation is a 
partial differential equation for $n$ functions 
$S^i=S^i(x^j, y^a)$  
\beq
\label{dwhj}
\der_i S^i + H(x^j, y^a, p_a^i= \der_a S^i) = 0.         
\eeq 
In order to approach a quantization based on the DW formulation 
we have to construct an analogue of the Poisson brackets, 
to identify the canonically conjugate variables, 
and to find the form of the equations of motion of dynamical 
variables in Poisson bracket formulation. 
Here a simplified sketch of the author's  recent 
approach to these questions 
in the case of scalar field theories is given 
(see (Kanatchikov, 1996) for more details). 
Unfortunately, 
no simple formula for the Poisson bracket is available so far,  
so that I have to present the whole construction 
which is a certain generalization of the 
well-known construction of  the Poisson bracket  
from the symplectic form in mechanics. 

Our starting point is what I call the 
{\em polysymplectic form} and denote 
$\Omega$. 
It generalizes to field theory 
the symplectic two-form known in mechanics 
and reduces to the latter at $n=1$.  
In local coordinates
\beq
\Omega:= -dy^a\we dp^i_a \we \om_i       ,
\eeq
where $\om:=dx^1\wedge...\wedge dx^n$ and $\om_i:=\der_i \inn \om$. 
Below the variables $z^v := (y^a, p_a^i)$ are called {\em vertical} 
and the variables $x^i$ {\em horizontal}. 
The polysymplectic form  
maps functions of the polymomentum phase space variables to vertical 
multivectors of degree $n$ and, 
more generally, horizontal $q$-forms,  
$\ff{q}:=\frac{1}{q!}
F_{i_1 ... i_q}(z,x) 
dx^{i_1}\wedge ...\wedge dx^{i_q}$,    
which play the role of dynamical 
variables   
to vertical multivectors of degree $(n-q)$,  
$\xx{n-q}{}:=  \frac{1}{(n-q)!} X^v{}\uind{i}{n-q-1}
\der_v\we \partial_{i_1}
\wedge...\wedge\partial_{i_{n-q-1}}$. Thus for all $0\leq q < n$
\beq
\label{themap}
\xx{n-q}_{\! \! \scriptsize \ff{q}}  
\inn\ \Om = d^V \ff{q} , 
\eeq
where 
$\dv \ff{q}:=\frac{1}{q!}\der_v F_{i_1 ... i_q}(z) dz^v \we 
dx^{i_1}\wedge ...\wedge dx^{i_q}$. For $q=n$ a similar map exists between 
$n$-forms $F\omega$ and vertical-vector-valued one-forms  
$\xt := \xt^v{}_k dx^k \otimes \der_v$: 
$\xt_{F\omega} \inn \Omega 
:=  \xt^v{}_k dx^k\we \der_v\inn \Omega 
= \dv (F\omega)$. 

The Poisson bracket of two forms of degree $r$ and $s$ 
for which the map (2.5) exists (those are called Hamiltonian)
is the following (Hamiltonian) $(r+s-n+1)$-form
\beq
\mbox{$\pbr{\ff{r}_1}{\ff{s}_2} := 
(-1)^{(n-r)}X_{1} \inn\ d^{V} \ff{s}_2
= (-1)^{(n-r)}X_{1} \inn\ X_{2} \inn\  \Omega $    } .
\eeq
This bracket obeys the axioms of a graded Lie algebra. In 
particular 
$$
\pbr{\ff{r}_1}{\ff{s}_2} 
= - (-)^{(n-r-1)(n-s-1)}\pbr{\ff{s}_2}{\ff{r}_1}. 
$$  
Moreover, it fulfills a generalized (graded and higher-order) 
Poisson property (see (Kanat\-chikov, 1996)).
Thus a generalization of  Lie and Poisson properties  
of the standard Poisson bracket, 
which 
	are known to 
underlie the standard canonical quantization, 
is obtained here. 
The question naturally arises as to whether this new 
algebraic structure 
on differential forms  
can be 
used as a starting point 
for a quantization  in field theory. 
Before addressing this question in the next section let us formulate 
the field equations of motion in terms of the bracket above.  
One can expect those are given by the bracket with $H$ or $H\omega$. 
In fact, introducing the total differential $\bd$ of a form:
$\bd \ff{p} := 
\der_iz^vdx^i\we\der_v\ff{p}+dx^i\we \der_i\ff{p}$, 
which  generalizes the total time derivative in mechanics, 
the equation of motion of a $p$-form dynamical variable 
assumes the form 
\beq
\mbox{{\bm $d$}}\ff{p}=
\pbr{H\om}{\ff{p}}
+d^{hor}\ff{p} , 
\eeq  
where $\pbr{H\om}{\ff{p}}
:=\xt_{H\om}\inn \dv \ff{p}$ 
and 
$d^{hor}\ff{p}:= dx^i\we \der_i\ff{p}$.  
To prove (2.7) substitute for the 
components of $\xt_{H\om}$ their values on extremals: 
$\xt^v{}_k = dz^v/dx^k$. 
The field equations in the $DW$ canonical form (2.2) are 
reproduced if, for instance, the forms $p_a^i\omega_i$ and $y^a$ 
are inserted into (2.7).

\section{ 
\large \bf 
QUANTIZATION 
AND 
GENERALIZED SCHR\"ODINGER EQUATION }

In order to develop  an  approach similar to 
the canonical quantization in 
the Schr\"odinger representation of quantum mechanics  
the pair(s) of the canonically conjugate variables have to be 
identified.  
Let us introduce  
$(n-1)$-forms associated with  
polymomenta: $p_a:=p_a^i\om_i$. 
Then 
the following set of  the canonical 
brackets in Lie subalgebra 
of $0$- and $(n-1)$-forms can be obtained from (2.5) and (2.6) 
\beq
\label{pbr1}
\pbr{p_a}{y^b}=\delta_a^b, 
\quad 
\pbr{p_a^i}{y^b \omega_j}=\delta^i_j \delta^b_a ,
\quad
\pbr{p_a}{y^b \omega_j}=\delta^b_a \omega_j . 
\eeq
We  quantize them using Dirac's correspondence rule 
that the graded Poisson bracket multiplied by $ i\hbar$ 
goes over into 
the graded commutator with 
the same symmetry property. 
In the ``$y$-representation'' 
the quantization of the 
first bracket in (3.1) yields 
\beq
\hat{p}_a = i\hbar \der_a . 
\eeq
Assuming  $\hat{p}{}^i_a= i\hbar \der_a \otimes \hat{p}{}^i$
and quantizing (3.1b) we obtain  
\beqa
[\hat{p}{}^i_a, \widehat{y^b \omega_j}]
&=& i\hbar \der_a \otimes \hat{p}{}^i \circ \widehat{y^b \omega_j}
- (-)^{(n-0-1)(n-(n-1)-1)} \widehat{y^b \omega_j} \circ
i\hbar \der_a \otimes \hat{p}{}^i
\nn \\
&=& i\hbar \delta _a^b \hat{p}{}^i \circ \widehat{ \omega_j}
+\hat{p}{}^i \circ \widehat{ \omega_j}\hat{y}{}^b i\hbar \der_a
- \widehat{ \omega_j}\circ\hat{p}{}^i \hat{y}{}^b i\hbar \der_a 
= i\hbar \delta^b_a \delta^i_j ,  
\nn 
\eeqa
whence 
\beq
\hat{p}{}^i \circ \widehat{ \omega}_j = \delta^i_j, 
\quad 
\hat{p}{}^i \circ \widehat{ \omega}_j
- \widehat{ \omega}_j\circ\hat{p}{}^i = 0  , 
\eeq
where $\circ$ denotes the composition of operators. 
To find a realization of (3.3) let us note 
that graded symmetry properties of the exterior product 
and of our  Poisson bracket 
should be incorporated in  
the algebraic system chosen for the representation. 
A natural, if not unique, choice seems to be the 
hypercomplex algebra of the space-time manifold 
(see e.g. Hestenes, 1966). 
Firstly, it reduces to the complex algebra in the 
case of quantum mechanics ($n=1$); secondly, 
it unifies the properties of operators 
$dx\we$ and $\der\inn$.   
On this basis we arrive to  
the  realization of (3.3) 
in terms of the hypercomplex imaginary units $\ga_i$ such that 
$\ga_i\ga_j+\ga_j\ga_i=g_{ij}$ 
($g_{ij}$ is the space-time metric tensor). One can take 
For instance, one can take 
\beq
\hat{p}{}^i = \kappa \ga^i \ga ,
\quad 
\widehat{ \omega}_j = \kappa^{-1} \ga \ga_j, 
\eeq 
where $\ga :=  i^{\frac{1}{2}n(n-1)} \sigma^{\half}
\ga_1 \ga_2 ... \ga_n$, 
$\sigma=$ sign(det$(g_{ij}))$, so that $\ga^2=1$.   
The quantity $\kappa$ of the dimension 
$[length^{-(n-1)}]$ appears in order to account for  
the physical dimensions of $p^i$ and $\omega_i$. 
The absolute value of its inverse is expected to be 
(formally) infinitesimal as $\omega_i$ is 
essentially  an infinitesimal volume element. 

The hypercomplex algebra of the space-time manifold 
appears here as a generalization of the complex algebra in 
the formalism of quantum mechanics. 
Therefore the wave function  
also can be  taken to be 
a hypercomplex-valued function  
on the configuration space of variables $(x^i,y^a)$,  
i.e. $\Psi = \psi I + \psi_\mu \ga^\mu + 
\psi_{\mu \nu} \ga^{[\mu} \ga^{\nu ]}
+ ... $  
An analogue of the 
Schr\"odinger equation for $\Psi$ 
can be obtained (guessed)  
from the requirements that   
(i) the DW HJ equation (2.3) would 
appear in the classical limit and 
(ii) the familiar quantum mechanical Schr\"odinger equation 
would be reproduced at $n=1$. 
Besides, the observation in Sect. 2 
that ``the DW Hamiltonian governs the exterior differential'' 
is essential.    
With these considerations in mind 
the following generalized Schr\"odinger equation can 
be put forward  
\beq
i\hbar \kappa \gamma^i \der_i \Psi = \what{H}\Psi ,
\eeq
where the quantity $\kappa$ appears again on dimensional 
grounds. The  left hand side is chosen to be the Dirac operator 
as it can be viewed at once  
as a multidimensional generalization of the partial time 
derivative 
and,  in a sense,  as an analogue of the 
exterior differentiation  
acting on hypercomplex functions. We show below that 
this equation  
does indeed reduce to the DW HJ equation in the 
quasiclassical limit, 
at least in the  case of scalar fields (see eq. (3.10)). 

Let us consider the interacting  
scalar field theory given by 
the  Lagrangi\-an density
$ L=\frac{1}{2} \der_i y^a \der^i y_a - V(y^a). $
For this system $p_i^a= \der_i y^a$ and the DW Hamiltonian function 
$ 
H =
\half  p^i_a p^a_i + V(y) .
$ 
In order to construct the operator $\widehat{H}$ a realization of 
$\widehat{p^i_a p^a_i}$ has to be found.  
From the  quantization of   the bracket 
$\pbr{\frac{1}{2} p^i_a p^a_i}{y^b \omega_j}= p^b_j$ 
we find 
$
\widehat{p^i_a p^a_i} = -\hbar^2 \kappa^2 \lapl ,
$
where  $\lapl:= \der_a \der^a$. Thus
\beq
\widehat{H}=-1/2\, \hbar^2 \kappa^2 \lapl + V(y). 
\eeq
To close the system of equations (3.5) it is sufficient 
to take 
$$
\Psi = \psi \, I + \psi^i \gamma_i   .
$$  
Then (3.5) 
reduces to the system of equations 
\beq
i\hbar \kappa \der_i \psi^i = \what{H} \psi   , 
\quad 
 i\hbar \kappa \der_i \psi = \what{H}  \psi_i 
\eeq 
which gives rise to the conservation law 
\beq
\der_i [\psib \psi^i + \psi \psib{}^i  ]
= \frac{i\hbar \kappa}{2} \der^a[\psib \derts_a \psi 
+ \psib^i \derts_a \psi_i ]    .       
\eeq
This could lead to a prescription for the calculation of 
the 
expectation values. However,  the corresponding 
scalar product is unlikely to stay positive definite. 
Moreover, 
the  obstacles of pure algebraic nature are known 
to a generalization of the quantum theoretical formalism 
to amplitudes  different from the real,  
quaternion or octonion valued functions 
(see e.g. (Adler, 1992)). 
A possible way out can be  
in 
replacing  
the hypercomplex wave function in (3.5) with a spinor one,  
while still keeping representing operators corresponding to forms 
in terms of the  hypercomplex numbers.   

  In order to consider 
 the quasiclassical limit of (3.5)  
let us take  the following 
hypercomplex generalization of the 
quasiclassical Ansatz for the wave function 
\beq
\Psi = R \, \exp (iS^\mu \ga_\mu / \hka) , 
\eeq
where the exponential function is defined via the series 
expansion. 
Substi\-tuting (3.9) into (3.5) and performing 
a slightly tedious calculation 
the result can be represented in the  form 
\beq 
\label{dwhjq}
\der_\mu S^\mu = -\half \der_a S^\mu \der_a S_\mu 
- V(y) 
+\half \hbar^2 \kappa^2  \frac{\lapl R }{R}  
\eeq 
which reproduces the DW HJ equation for scalar fields 
(cf. eq. (2.3)) in the quasiclassical limit 
$\hbar \kappa \rightarrow 0$. 
However, 
besides the DWHJ equation 
two other conditions on the ``HJ functions'' $S^i$: 
$\der_i S^i = \frac{S^i}{|S|} \der_i |S|$ 
and  
$\der_a S^i \der^a S_i = \der_i S^i $,    
where $|S|:=\sqrt{S_iS^i}$, 
arise here, 
so that no {\em complete} DWHJ theory is 
reproduced.   
It is interesting to note  
that the ``quantum potential'' term in (3.10),   
$\hbar^2 \kappa^2 \lapl R / R $,  
is of the similar form as  
in quantum mechanics (cf. Bohm e.a., 1987). 
 
For a free real scalar field 
$\what{H} = -\half \hbar^2 \kappa^2 \der_{yy} + 
\half \frac{m^2}{\hbar^2} y^2 , 
$
and eq. (3.5) can be solved by the separation of variables:
$\Psi(x,y)=\Phi(x)f(y)$, 
where $\Phi(x):=\phi(x)+\phi_i(x)\ga^i$ 
and  $f(y)$ is a function. 
This leads to the eigenvalue problem for the DW Hamiltonian operator: 
$\what{H} f = \chi f$. In the present case it is just the 
harmonic oscillator problem in the space of field 
variables, so that the eigenvalues of $\what{H}$ are 
$\chi_{{}_N}=\kappa m (N+1/2)=:\kappa m_N $ 
and the eigenfunctions are those of the harmonic oscillator. 
The scalar part of $\Phi$ satisfies the 
Klein-Gordon equation 
$\Box \phi = -\chi^2 / \hbar^2 \kappa^2 \phi$,
and the vector part obeys 
$ \phi_i=\frac{i\hbar \kappa}{\chi} \der_i \phi $.  
Let us note that the quantity $\kappa$ 
cancels out in 
the equations  governing the space-time behavior of $\Psi$. 
Now, for the ground state, $N=0$, the scalar part 
of $\Psi$ assumes the form 
\beq
\psi_{0,\bk }(y, \bx , t )
\sim e^{i \omega_{0,\bk } t - i \bk \cdot \bx} 
e^{-\frac{m}{2 \kappa \hbar^2} y^2} , 
\eeq
where 
$ \omega_{0,\bk }^2 - \bk^2
= (\half\frac{m}{\hbar})^2 $ and a normalization factor is omitted.  
In general, any solution of eqs. (3.7) is a linear combination of 
$\Psi_{N,\bk}$.

Ignoring the problem with the positive indefiniteness of the scalar 
product implied by (3.8) 
our hypercomplex wave function $\Psi(y,x)$  
can be interpreted 
as the probability 
amplitude of finding the value $y$ of the field in 
the infinitesimal vicinity of the 
space-time point $x$. 
If so, we can try to relate it to the Schr\"odinger 
wave functional in quantum field theory  $\Psi_{}([y(\bx)],t)$  
which is the probability amplitude of finding a field configuration 
$y(\bx)$ on a space-like hypersurface of constant time $t$. 

For instance,  for a free scalar field 
the Schr\"odinger  vacuum state 
functional 
\beq
\Psi_{ 0}([y(\bx )], t)
=\eta\ \exp \left( \frac{i}{\hbar} E_0 t 
-\frac{1}{\hbar} \int \! 
d \bx \, y(\bx ) 
\half (-\nabla^2_{\bx } + m^2/ \hbar^2)^{\half } y(\bx )  
\right) 
\eeq 
can be expressed  as the infinite product of the harmonic oscillator 
wave functions over all points in the $\bk$-space 
(cf. e.g. Hatfield, 1992) 
\beq 
\lim_{V\rightarrow\infty}
\eta\ 
\prod_{\bk }
\exp \, \half \left(  i  \omk t 
-  \frac{1}{V \hbar } \ \omk \tilde{y}{}^2(\bk)  \right), 
\eeq 
where $\omk := ( m^2/ \hbar^2 + \bk^2)^\half$, 
$E_0 = 
\half \hbar \int_V \! d\bx \int \! \frac{ d\bk}{(2\pi)^{n-1}}  \omk 
$ is the (divergent) vacuum state energy, 
$V$ is an ``infinitely large''  volume element, 
$\eta$ is a normalization factor,
and the Fourier series expansion 
$y(\bx ) = 
\frac{1}{V} \sum_\bk \tilde{y}{}(\bk)  
e^{i\bk \bx } $ 
is used 
in passing  from (3.12) to (3.13).

At the same time the amplitude of finding the  configuration 
$y(\bx)$ of 
the field can be composed from the infinite set of the 
amplitudes 
of finding the corresponding values $y=y_{\bx}$ of the field 
in the points $\bx$ of the equal-time hypersurface.  
These amplitudes are given by  our wave function 
$\Psi(y=y_{\bx},x=(\bx, t))$.  
Now, if the correlations between the values of 
the field in space-like separated points 
are neglected, 
the composed vacuum state amplitude can be written as 
the infinite product over all points of the $\bx$-space 
of the lowest eigenvalue solutions  (3.11), 
that is   
$$ 
\prod_{\bk }  
e^{i  \omega_{0,\bk} t -i \bk \cdot \bx} 
\prod_{\bx \in V }
\exp \left( -\frac{m}{2 \kappa \hbar^2} y_{\bx }^2 \right) 
,
$$ 
where the product over  $\bk$ 
accounts for filling of all $\bk$-states in the vacuum state.  
Inserting the formal Fourier series expansion for $y_\bx$ 
and using a discretization in both $\bk$-  and $\bx$- space and 
the identity 
$\prod_{\bk} e^{i\bk \bx} = 1$ 
this expression can be transformed to the form similar to (3.13)
\beq
\lim_{V\rightarrow\infty \atop Q\rightarrow\infty} 
\prod_{\bk }
\exp \,  \left(  i  \omega_{0,\bk} t 
- \frac{Q}{2 (2\pi )^{n-1} V \hbar^2 \kappa} \ m\ \tilde{y}{}^2(\bk)  
\right), 
\eeq
where $Q:=\int_Q d^{n-1} \bk$ plays the role of the ultra-violet cutoff. 

The amplitude in (3.14) is different from 
that in (3.13) in two respects. 
Firstly, in the second term of (3.14) 
we have obtained the proper mass $m$ 
instead of  the frequency $\omega_\bk$ in (3.13).   
This is probably a result of our above neglect of 
space-like correlations which appear in the 
standard theory due to the non-local character of the 
operator $\sqrt{m^2/\hbar^2 - \nabla_\bx^2}$ 
in (3.12),    
and which are 
accounted for in the part of 
the Feynman propagator 
non-vanishing at space-like separations.   
However, the study of Green  functions  
of the 
second-order consequence of 
(3.7): 
$ \hbar^2 \Box \psi = - \frac{1}{\kappa^2} \what{H}{}^2 \psi$,  
demonstrates that 
these  correlations are not 
beyond the scope of the present approach, so that 
the neglect of space-like correlations is of a technical character.  
Secondly, we face the problem of a proper interpretation of the 
parameter $\kappa$.  The latter appeared in 
(3.4) as essentially the inverse of 
an  infinitesimal $(n-1)$-volume element.  
It can, therefore,  naturally be identified 
with the ultra-violet cutoff scale $Q$ divided by $(2\pi )^{n-1}$.  
With such an identification 
eq. (3.14) provides us with a very long wave 
$(|\bk| \ll  m)$ limit 
(in which the space-like correlations are vanishing) 
of the Schr\"odinger vacuum state 
functional (3.13). 
Therefore, the composed amplitude in (3.14) 
appears to be consistent with the 
standard result in (3.12) within 
the simplifying rough approximation of neglect of 
non-vanishing part of the Feynman propagator at 
space-like separations.

Summarizing, we have argued  that a 
quantization of field theory 
based on the polymomentum Hamiltonian formulation 
of De Donder-Weyl 
leads to an interesting hypercomplex 
generalization of the formalism of quantum theory  
which is different from the previously 
considered versions of the quaternionic quantum mechanics 
and related approaches (cf. Adler, 1995).  
However, serious efforts are still required for understanding 
of 
how  the present approach to quantization  
can be related to 
or complement 
the modern notion of the quantum field. \\

\medskip

{\normalsize \bf ACKNOWLEDGMENTS} 

\medskip

I would like to thank Prof. K.-E. Hellwig 
for a kindly offered to me opportunity 
to present this work at the 
QS96 conference in Berlin.  \\

\bigskip 

{\normalsize \bf REFERENCES}\\
\medskip \\
\renewcommand{\baselinestretch}{0.95} 
\small
Adler, S.L.   (1995). {\em Quaternionic Quantum Mechanics and 
Quantum Fields}, 
Oxford Univ. Press, New York. \\
Bohm, D., Hiley, B.J. and Kaloyerou, P.N. (1987). 
{\em Physics Reports}, 
{\bf 144}, 349 \\
Born, M. (1934). {\em Proceedings of the Royal Society (London)}, 
{\bf A143}, 410. \\
Hatfield, B.  (1992). {\em Quantum Field Theory of Point 
Particles and Strings}, Addison-Wesley, Redwood City. \\
Hestenes, D. (1966). {\em Space-time algebra}, Gordon and Breach, New York \\
Kanatchikov, I.V. (1996). {\em Canonical Structure of Classical 
Field Theory in the Polymomentum Phase Space}, 
to appear in {\em Reports on Mathematical Physics, } 
{\bf 41} No. 1, hep-th/9709229. \\ 
Kastrup, H.A. (1983). {\em Physics Reports}, {\bf 101} 1 \\
Rund, H. (1966). {\em The Hamilton-Jacobi Theory in the Calculus of 
Variations}, D. van Nostrand, Toronto.  \\
Volterra, V. (1959). {\em Theory of Functionals and Integral and 
Integro-Differential Equations}, Dover Publ.,  New York.\\ 
Weyl, H. (1934). {\em Physical Review}, {\bf 46}, 505. 

\end{document}